\documentclass[12pt,english]{paper}
\usepackage{amssymb}
\usepackage[T1]{fontenc}
\usepackage[latin1]{inputenc}
\usepackage{graphicx}
\usepackage{cite}
\makeatletter

\makeatletter

\makeatletter

\usepackage{geometry}

\makeatletter

\usepackage{epsfig}

\makeatother

\makeatother

\makeatother

\usepackage{babel}
\makeatother
\begin{document}

\title{Search for Ultra-High Energy WIMPs by detecting neutrino signatures from the earth core}

\author{Ye Xu}

\maketitle

\begin{flushleft}
 School of Information Science and Engineering,  Fujian University of Technology, Fuzhou 350118, China
\par
e-mail address: xuy@fjut.edu.cn
\end{flushleft}

\begin{abstract}
I study the possibility of probing Ultra-High Energy (UHE from now on) dark matter particles $\chi$ (E$_{\chi}$> 10$^8$ GeV) due to the decay of superheavy dark matter via detection of UHE neutrino signatures from the earth core. The UHE WIMP event rates are estimated at the energies of 100 PeV, 1 EeV, 10 EeV, 100 EeV and 1 ZeV in JEM-EUSO and at the energy of 1 ZeV in IceCube, respectively. The diffuse neutrino contaminations are also roughly estimated in JEM-EUSO and IceCube. I found that the extreme energy WIMPs can be detected by JEM-EUSO above 1 EeV and by IceCube above 1 ZeV in the $\chi$ mass range of 10 GeV - 140 GeV.
\end{abstract}


\section{Introduction}
It is indicated by the Planck data with measurements of the cosmic microwave background
 that $26.6\%$ of the overall energy density of
the Universe is non-baryonic dark matter \cite{Planck2015}. Weakly Interacting Massive Particles (WIMPs from now on),
predicted by extensions of the Standard Model of particle physics,
are a class of candidates for dark matter\cite{GDJ}. They are
distributed in a halo surrounding a galaxy. A WIMP halo of a galaxy
with a local density of 0.3 GeV/cm$^3$ is assumed and its relative
speed to the Sun is 230 km/s\cite{JP}. At present, one mainly searches for thermal
WIMPs via direct and indirect detections\cite{CDMSII,CDEX,XENON1T,LUX,PANDAX,AMS-02,DAMPE,fermi}. Because of the very small cross sections of the interactions
between these WIMPs and nucleus (maybe O(10$^{-47}$ cm$^2$))\cite{XENON1T,PANDAX}, so far one has not found dark matter yet.
\par
It is a reasonable assumption that there exist various dark matter particles in the Universe. Then it is possible that
this sector may comprise of non-thermal (and non-relativistic) components. And these particles may also contain a small
component which is relativistic and highly energetic. Although the fraction of these relativistic dark matter particles
is small in the Universe, their large interaction cross sections (including between themselves and between them and the
Standard Model (SM from now on) particles) make it possible to find them. Due to the reasons mentioned above, one has
to shift more attention to direct and indirect detection of UHE WIMPs. In fact, A. Bhattecharya, R. Gaudhi and A. Gupta have discussed the possibility that the PeV events are UHE dark matter particles at IceCube in their work\cite{BGA}.
\par
The relativistic WIMPs are mainly generated by two mechanisms in the Universe. one is through the collision of UHE cosmic ray particles and
thermal WIMPs. This collision will result in some UHE WIMPs flux. The other is that UHE WIMPs can also originate from
the early Universe. There are a non-thermal dark sector generated by the early Universe with its bulk comprised of a very massive
relic $\phi$ in the Universe. This superheavy dark matter decays to another much lighter WIMPs $\chi$ and its lifetime is greater
than the age of the Universe. This lead to a small but significant flux of UHE WIMPs\cite{LT, BGA, EIP, BLS, BKMTZ}.
\par
In the present work, I only focus on detection of the UHE WIMPs $\chi$ induced by the decay of superheavy dark matter $\phi$ ($\phi\to\chi\bar{\chi}$).
In what follows, we will discuss the possibility of probing UHE WIMPs due to the decay of superheavy dark matter via detecting
UHE neutrino signatures from the earth core. We make an assumption that WIMPs only annihilate with each other via $\chi\bar{\chi}\to\nu\bar{\nu}$. And UHE neutrinos are generated by the annihilation between the UHE WIMPs and thermal
ones captured by the Earth and accumulated in the earth core. These neutrinos can be detected by km$^3$ neutrino telescopes, such as IceCube\cite{icecube}, ANTARES\cite{antares}, KM3NeT\cite{KM3NeT}, and satellite detectors, such as JEM-EUSO\cite{JEM-EUSO}.
\par
We will estimate those neutrino event rates and the background from diffuse neutrinos in IceCube and JEM-EUSO in the present paper.
\section{UHE WIMPs flux in the Galaxy}
We consider a scenario where the dark matter sector is composed of at least two particle species in the Universe. One is a co-moving non-relativistic scalar species $\phi$, with mass $m_{\phi} \geq$ 100 PeV, the other is much lighter particle species $\chi$ ($m_{\chi} \ll m_{\phi}$), due to the decay of $\phi$, with a very large lifetime. $\phi$ does not have any decay channels to SM particles. And $\chi$ comprises the bulk of present-day dark matter.
\par
The lifetime for the decay of heavy dark matter to SM particles is strongly constrained ($\tau \geq 10^{27}-10^{28}s$) by diffuse gamma and neutrino observations\cite{MB,RKP}. However, since the superheavy dark matter $\phi$ does not decay to SM particles in my work, the constraints on the $\phi$ lifetime are only those based on cosmology\cite{IOT,NS,DMQ,popolo}. Considering these constraints and the present-day relic abundance, I assume that the lifetime of $\phi$ is greater than the age of the Universe, i.e., $\tau_{\phi}\geq10^{17}$s\cite{BGA}.
\par
The UHE WIMPs flux in the Galaxy is obtained via the following equation\cite{BLS}:
\begin{center}
\begin{equation}
\Phi_{\chi}=1.7\times10^{-12}cm^{-2}s^{-1}\times\frac{10^{28}s}{\tau_{\phi}}\times\frac{1PeV}{m_{\phi}}.
\end{equation}
\end{center}
$\tau_{\phi} $ is taken to $10^{18}$s in the present paper.
\section{WIMPs accumulation in the Earth}
When the WIMPs wind sweeps through the Earth, WIMPs could collide with the  matter in the Earth and lose their kinetic energy. Then these WIMPs can be captured by the Earth's gravity and enter the earth core. After a long period of accumulation, the WIMPs inside the Earth can begin to annihilate into neutrinos at an appreciable rate. The impact of self-capture for the Earth is suppressed by the fact that the escape velocity is low compared to typical galactic WIMP velocities, and so the typical WIMP self-scattering simply replace one captured WIMP with another\cite{ARZ}. So the self capture from WIMP-WIMP self-interaction is ignored in the present paper. The number N of WIMPs in the Earth obeys the equation\cite{BCH}
\par
\begin{center}
\begin{equation}
\frac{dN}{dt}=C_{cap}-2\Gamma_{ann}-C_{evp}N
\end{equation}
\end{center}
\par
where $C_{cap}$ is the capture rate, $\Gamma_{ann}$ is the annihilation rate and $C_{evp}$ is the evaporation rate.
\par
\begin{center}
\begin{equation}
\Gamma_{ann}=\frac{1}{2}\int d^3xn^2(\overrightarrow{x})\left\langle\sigma_{ann}\upsilon\right\rangle=\frac{1}{2}C_{ann}N^{2}(t)
\end{equation}
\end{center}
\par
where $\left\langle\sigma_{ann}\upsilon\right\rangle$ is the WIMP annihilation cross section for$\chi\bar{\chi} \to \nu\bar{\nu}$ and $n(\overrightarrow{x})$ is the number density of WIMPs at position $\overrightarrow{x}$ inside the Earth, such that the total number of WIMPs in the Earth is $N=\int d^3xn(\overrightarrow{x})$. After capture, subsequent scatterings thermalize the WIMPs to the earth temperature $T_e$, such that their density $n(\overrightarrow{x})$ acquires the spherically symmetrical Boltzmann form
\par
\begin{center}
\begin{equation}
n(r)=n_0exp[-m_{\chi}\phi(r)/T_e]
\end{equation}
\end{center}
\par
where $n_0$ is the central WIMP number density, $m_{\chi}$ is the WIMP mass and $\phi(r)$ is the Newtonian gravitational potential inside the Earth. The WIMP distribution in the Earth is obtained by\cite{BCH,FSP},
\par
\begin{center}
\begin{equation}
n(r)=n_0exp\left(-\frac{r^2}{r_{\chi}^2}\right),  \quad with\quad r_{\chi}=\left(\frac{3T_e}{2\pi G_N\rho_em_{\chi}}\right)^{1/2}\approx150km\sqrt{\frac{TeV}{m_{\chi}}}
\end{equation}
\end{center}
\par
where $\rho_e\approx$ 13 g/cm$^3$ and $T_e\approx$ 5700 K are the matter density and temperature at the Earth center, respectively. One finds that WIMPs are concentrated around the center of the Earth. The evaporation rate is only relevant when the WIMP mass m$_{\chi}$<5 GeV\cite{gould,KSW,nauenberg,GS}, which are lower than our interested mass scale (m$_{\chi}$ $\geq$ 10 GeV). Thus the WIMPs evaporation is ignored in my work. By solving Eq.(2), we obtain\cite{GS}
\par
\begin{center}
\begin{equation}
\Gamma_{ann}=\frac{C_{cap}}{2}tanh^2\left(\frac{t}{\tau}\right)
\end{equation}
\end{center}
\par
where $\tau=(C_{cap}C_{ann})^{-\frac{1}{2}}$ is the time scale when the WIMPs capture and annihilation in the earth core reaches the equilibrium state. Taking $t$ (about 10$^{17}$s) the age of the Earth, we have\cite{LLL}
\par
\begin{center}
\begin{equation}
\frac{t}{\tau}\approx1.9\times10^4\left(\frac{C_{cap}}{s^{-1}}\right)^{1/2}\left(\frac{<\sigma_{ann}\upsilon>}{cm^3s^{-1}}\right)^{1/2}\left(\frac{m_{\chi}}{10 GeV}\right)^{3/4}
\end{equation}
\end{center}
\par
where $C_{cap}$ is proportional to $\displaystyle\frac{\sigma_{\chi N}}{m_{\chi}}$\cite{JKK} and  $\sigma_{\chi N}$ is the scattering cross section between the nucleons and WIMPs. The $C_{cap}$ calculation in Ref.\cite{LE} is adopted but $\sigma_{\chi N}$ is taken to be 10$^{-47}$ cm$^2$\cite{XENON1T,PANDAX}.
\par

\section{Annihilation cross section between thermal and ultra-relativistic WIMPs}
Mirco Cannoni studied relativistic and non-relativistic annihilation of dark matter using an effective field theory approach in Ref.\cite{cannoni}. This approach is used to estimate the annihilation cross section between thermal and ultra-relativistic WIMPs in my work.
\par
A dark matter fermion field $\chi$ couples to other fermion field $\psi$ through an effective dimension-6 operator of the type
\par
\begin{center}
\begin{equation}
\mathcal{L}=\frac{\lambda_a\lambda_b}{\Lambda^2}(\bar{\chi}\Gamma_a\chi)(\bar{\psi}\Gamma_b\psi)
\end{equation}
\end{center}
\par
where $\psi$ are the SM fermions, $\lambda_{a,b}$ are dimensionless coupling associated with the interactions described by combination of Dirac matrices $\Gamma_{a,b}$ and $\Lambda$ is the energy scale below which the effective field theory is valid. compared to $\Lambda$ and $m_{\chi}$, the $\psi$ masses can be neglected. In a word, the $s$ and $t$ channel annihilations are reduced to the effective vertex corresponding to the lagrangian Eq. (8) (see Fig. 1 in Ref.\cite{cannoni}).
\par
To get the formula for the cross section of the WIMP annihilation $\left\langle\sigma\upsilon\right\rangle$ in a useful form, it is convenient to define the reduced cross section
\par
\begin{center}
\begin{equation}
\sigma_0=\frac{1}{2m_{\chi}^2}\frac{1}{32\pi}\omega
\end{equation}
\end{center}
\par
and the effective cross section
\par
\begin{center}
\begin{equation}
\sigma_{\Lambda}=\frac{\lambda_{a}^2\lambda_{b}^2}{4\pi}\frac{m_{\chi}^2}{\Lambda^4}
\end{equation}
\end{center}
\par
where $\omega$ is an integrated squared matrix element summed over the final spins and averaged over the initial spins. the reduced cross section Eq. (9) becomes
\par
\begin{center}
\begin{equation}
\sigma_0=\sigma_{\Lambda}\left(a_2y^2+\frac{a_1}{4}y+\frac{a_0}{16}\right)
\end{equation}
\end{center}
\par
where $a_0, a_1, a_2$ are pure numbers and $\displaystyle y=\frac{s}{4m_{\chi}^2}$. $s$ is the Mandelstam invariant $s=(p_1+p_2)^2$, where $p_{1,2}$ are the four momenta.
\par
In the non-relativistic limit, $\displaystyle x=\frac{m_{\chi}}{T_{\chi}}\gg1$, where $T$ is the kinetic energy,
\par
\begin{center}
\begin{equation}
\left\langle\sigma\upsilon\right\rangle_{nr}\approx\sigma_{\Lambda}a_2\left(1+\frac{k}{4}-\frac{3k}{8x}\right)
\end{equation}
\end{center}
\par
where $\left\langle\sigma\upsilon\right\rangle_{nr}$ is the non-relativistic cross section and $\displaystyle k=\frac{a_1}{a_2}$. The different k values correspond to the different WIMP annihilation models. The chiral annihilation is only considered in my work, in other words, k is taken to be -2\cite{cannoni}.
\par
In the ultra-relativistic limit, $x\ll1$,
\par
\begin{center}
\begin{equation}
\left\langle\sigma\upsilon\right\rangle_{ur}\approx\sigma_{\Lambda}a_2\frac{3}{x^2}
\end{equation}
\end{center}
\par
where $\left\langle\sigma\upsilon\right\rangle_{ur}$ is the ultra-relativistic cross section.
\par
The WIMPs in dark mater halo are thermal and non-relativistic ($x_{nr}\gg1$) and the ones from the decay of superheavy dark matter are ultra-relativistic ($x_{ur}\ll1$). Then we can obtain
\par
\begin{center}
\begin{equation}
\frac{\left\langle\sigma\upsilon\right\rangle_{ur}}{\left\langle\sigma\upsilon\right\rangle_{nr}}\approx\frac{\displaystyle\frac{3}{x_{ur}^2}}{1+\displaystyle\frac{k}{4}-\displaystyle\frac{3k}{8x_{nr}}},
\end{equation}
\end{center}
\par
where $\left\langle\sigma\upsilon\right\rangle_{nr} \sim 3\times10^{-26}cm^3s^{-1}$ for WIMPs that are a thermal relic\cite{BHS}.
\section{Neutrino interaction length in the Earth}
Since UHE neutrinos due to the annihilation between WIMPs are from the earth core, the neutrino interaction length in the Earth, $L_{\nu}$, should be the same order of magnitude as the earth radius, $R_e$, for detecting them. $L_{\nu}$ can be calculated through the scattering cross section between neutrinos and matter inside the Earth,
\par
\begin{center}
\begin{equation}
L_{\nu}=\frac{1}{N_A\rho\sigma_{\nu N}}
\end{equation}
\end{center}
\par
where $N_A$ is the Avogadro constant, $\rho$=5.5 g/cm$^3$ is the density of the Earth and $\sigma_{\nu N}$ is the scattering cross section between neutrinos and matter inside the Earth. For neutrino energies above $10^{15}$ eV, good representations of the cross section are given by simple power-law forms\cite{GQRS}:

\begin{center}
\begin{equation}
\sigma_{CC}({\nu}N)=2.69\times10^{-36} cm^2 \left(\frac{E_{\nu}}{1 GeV}\right)^{0.402}
\end{equation}
\end{center}
\begin{center}
\begin{equation}
\sigma_{NC}({\nu}N)=1.06\times10^{-36} cm^2 \left(\frac{E_{\nu}}{1 GeV}\right)^{0.408}
\end{equation}
\end{center}
\begin{center}
\begin{equation}
\sigma_{CC}(\bar{\nu}N)=2.53\times10^{-36} cm^2 \left(\frac{E_{\nu}}{1 GeV}\right)^{0.404}
\end{equation}
\end{center}
\begin{center}
\begin{equation}
\sigma_{nC}(\bar{\nu}N)=0.98\times10^{-36} cm^2 \left(\frac{E_{\nu}}{1 GeV}\right)^{0.410}
\end{equation}
\end{center}

\section{Results}
UHE WIMPs reach the earth core and annihilate with WIMPs captured by the Earth. The UHE neutrinos generated by this annihilation pass through the Earth and interact with ice or air on the Earth. Compared to UHE WIMPs, the velocities of WIMPs captured in the earth core can be ignored. These neutrinos are along the same direction as UHE WIMPs. So the WIMP events from the far side of the Earth and through the earth core can only be detected by the km$^3$ neutrino and satellite detectors. The number of these UHE neutrinos detected by a detector, $N_{det}$, is
\par
\begin{center}
\begin{equation}
N_{det}=R\times T\times \int n(\overrightarrow{x})\left\langle\sigma\upsilon\right\rangle_{ur}\frac{\Phi_{\chi}}{\upsilon}\frac{2\pi {R_e}^2sin\theta}{4{\pi}{R_e}^2}P d\theta d^3x
\end{equation}
\end{center}
\par
where $\theta$ is the polar angle of neutrinos from the earth core, R is the duty cycle for an experiment, $\upsilon$ is the velocity of a UHE WIMP and T is the live time of an experiment (T is ten years in my work). $P=exp\left(-\displaystyle\frac{R_e}{L_{earth}}\right)\left[1-exp\left(-\displaystyle\frac{D}{L_{ice,air}}\right)\right]$ is the probability that the UHE neutrino interacts with ice or air after traveling a distance between $R_e$ and $R_e+D$, where D is the effective depth of the detector and $L_{earth,ice,air}$ are the neutrino interaction lengths with the Earth, ice and air, respectively. Since WIMPs captured by the Earth are concentrated on the earth core, UHE WIMP signatures should be from the earth core (a radius of 1500 km)
\subsection{JEM-EUSO}
JEM-EUSO is a space science observatory to explore the extreme-energy cosmic rays and upward neutrinos in the Universe\cite{eusomission}. It will place on International Space Station whose altitude is about 400 km, and orbit the earth every $\sim$ 90 min. The JEM-EUSO telescope has a wide field of view (FOV: $\pm30^{\circ}$) and observes extreme energy particles in the two modes (nadir and tilted modes) via fluorescent and Cherenkov photons due to the development of extensive air showers. I only consider the nadir mode of JEM-EUSO in my work. JEM-EUSO has a observational area of $2\times10^5$ km$^2$ in this mode. The duty cycle for JEM-EUSO, R, is taken to 10\%.
\par
Fig.1-5 show that the different detected regions in the ($m_{\chi}$,$E_{nu}$) plane are estimated at the different $\phi$ masses ($m_{\phi}$=200 PeV, 2 EeV, 20 EeV, 200 EeV and 2 ZeV)in JEM-EUSO, respectively. The different colors represent the different number ranges of neutrino signatures in ten years ($N_{det}$) and these regions are grouped in the $\chi$ mass range of 10 GeV - 140 GeV. We can see that there are four peaks for $N_{det}$ at the $\chi$ mass of about 14 GeV, 23 GeV, 26 GeV and 52 GeV in these figures since there are four peaks in the capture rate for WIMPs at the relevant masses. From these figures, we can conclude that the number of neutrino signatures $N_{det}$ increases with the $\phi$ mass, $m_{\phi}$, since $N_{det}$ are proportional to the $\chi$ energy $E_{\chi}$ and $E_{\chi}=\displaystyle\frac{1}{2}m_{\phi}$. The background due to diffuse neutrinos in JEM-EUSO is roughly estimated with a diffuse neutrino flux of $E^2\Phi_{\nu}=0.95\pm0.3\times10^{-8}GeV cm^{-2}s^{-1}sr^{-1}$\cite{BGA}, where $\Phi_{\nu}$ represents the per-flavor flux, in the present paper (see Fig.6). We can see that the diffuse neutrino contamination level decreases to O(1) events per ten years above the energy of about 1.1 PeV, O(0.1) events per ten years above the energy of about 2.5 PeV and O(0.01) events per ten years above the energy of about 5 PeV in Fig.6. These figures show that the neutrinos from ultra-relativistic WIMPs annihilation in the earth core can be detected by JEM-EUSO in the energy range of O(1PeV) - O(10PeV). Fig.1-5 show the maximum UHE WIMP event rates are O(0.1) events per ten years at $m_\phi$ = 200 PeV,  O(1) events per ten years at $m_\phi$ = 2 EeV, O(10) events per ten years at $m_\phi$ = 20 EeV, O(100) events per ten years at $m_\phi$ = 200 EeV and O(10$^3$) events per ten years at $m_\phi$ = 2 ZeV above the neutrino energy of 2.5 PeV, respectively.
\subsection{IceCube}
IceCube is a km$^3$ neutrino telescope and can detect three flavour neutrinos via detecting the secondary particles, that in turn emit Cherenkov photons, produced by the interaction between neutrinos and the Antarctic ice. In the present paper, I assume that all UHE neutrinos detected by IceCube interact with the ice within its volume and take $A_{eff}$ $\approx 1 km^2$, D $\approx$ 1 km and R $\approx$ 100\%.
\par
Fig.7 show the detected region in the ($m_{\chi}$,$E_{\nu}$) plane is estimated at $m_{\phi}$=2 ZeV in IceCube. This region is grouped in the $\chi$ mass range of 10 GeV - 65 GeV. The background from diffuse neutrinos in IceCube is roughly estimated with the same neutrino flux as that of JEM-EUSO (see Fig.8). We can see that the diffuse neutrino contamination level decreases to O(0.01) events per ten years above the energy of about 130 TeV and O(0.1) events per ten years above the energy of about 70 TeV in Fig.8. These figures show that the neutrinos from ultra-relativistic WIMPs annihilation in the earth core can be detected by IceCube in the energy range of 70 TeV - 3 PeV. Fig.7 shows the maximum UHE WIMP event rate is $\sim$ 2 events per ten years at the neutrino energy of about 2.2 PeV when $m_{\phi}$ = 2 ZeV.

\section{Conclusion}
Here I assume the requirements of detectable region in the ($m_{\chi}$,$E_{\nu}$) plane are that at least one UHE WIMP event are detected by the relevant detector in ten years and its signal to background ratio has to be greater than 10. The UHE neutrinos, which are from the earth core, can be detected by satellite telescopes and km$^3$ neutrino detectors under ice or water. So it is possible that one searches for UHE WIMPs via the annihilation channel of $\chi\bar{\chi}\to\nu\bar{\nu}$. I found that UHE WIMPs can only be detected in JEM-EUSO when $E_{\chi}\gtrsim1 EeV$ ($m_{\phi}\gtrsim2 EeV$) and in IceCube when $E_{\chi}\gtrsim1 ZeV$ ($m_{\phi}\gtrsim2 ZeV$).
\par
Since $\Phi_{\chi}$ is proportional to $\displaystyle\frac{1}{\tau_{\phi}}$, the above conclusion is actually depended on the lifetime of superheavy dark matter. The energy threshold will be enhanced by the longer $\tau_{\phi}$ in this indirect detection of UHE WIMPs.
\section{Acknowledgements}
This work is supported in part by the Natural Science Foundation of
Fujian Province in China under the contract No. 2015J01577 and science fund of Fujian
University of Technology under contract No. GY-Z13114.
\par

\newpage

\begin{figure}
 \centering
 \includegraphics[width=0.9\textwidth]{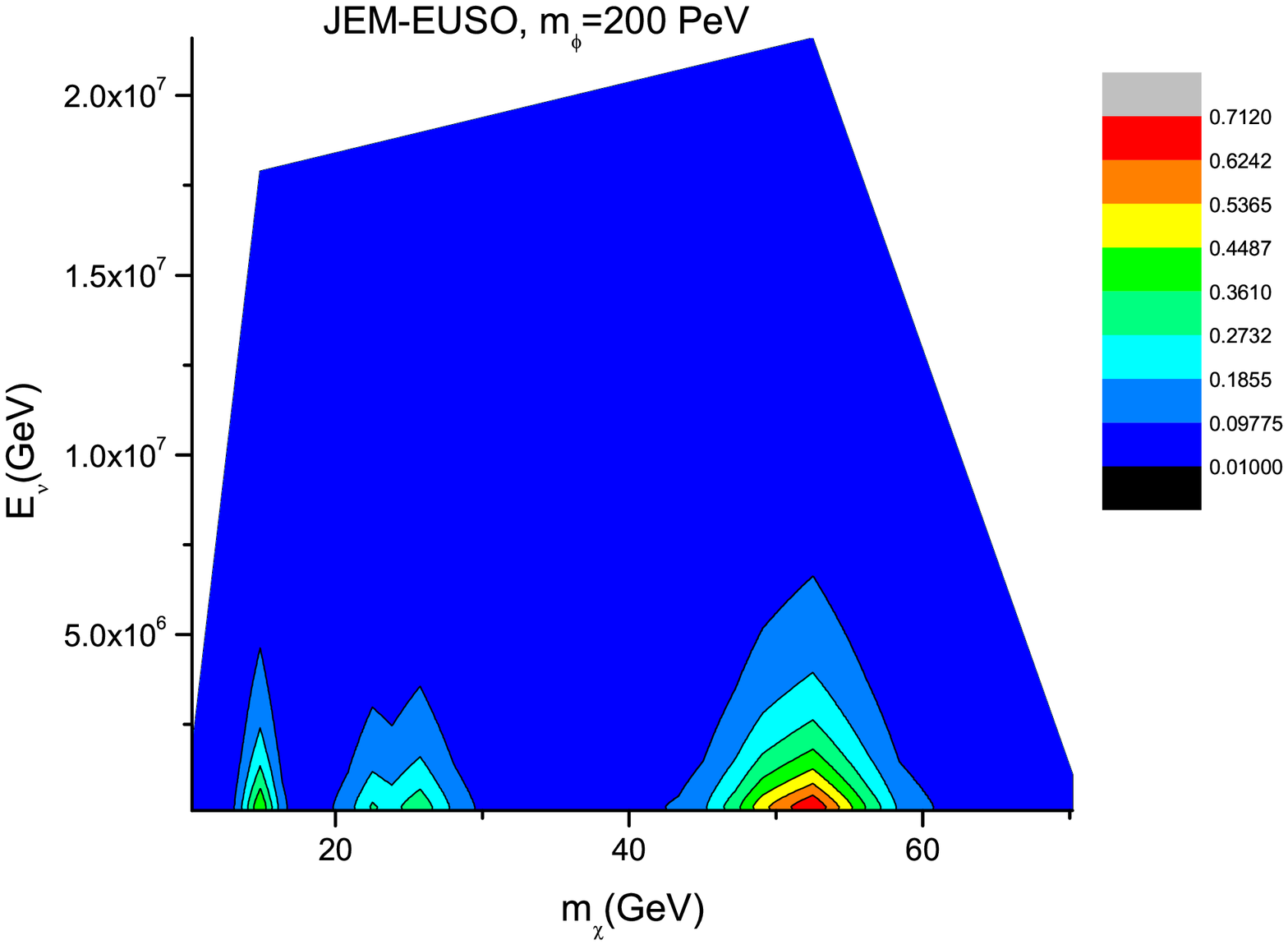}
 \caption{The detected region in the ($m_{\chi}$,$E_{\nu}$) plane is estimated at $m_{phi}=200 PeV$ in JEM-EUSO. $E_{\nu}$ is the energies of neutrinos due to the UHE WIMPs annihilation. The colors represent the number of neutrino signatures from UHE WIMPs in ten years.}
 \label{fig:EUSO_200PeV}
\end{figure}

\begin{figure}
 \centering
 \includegraphics[width=0.9\textwidth]{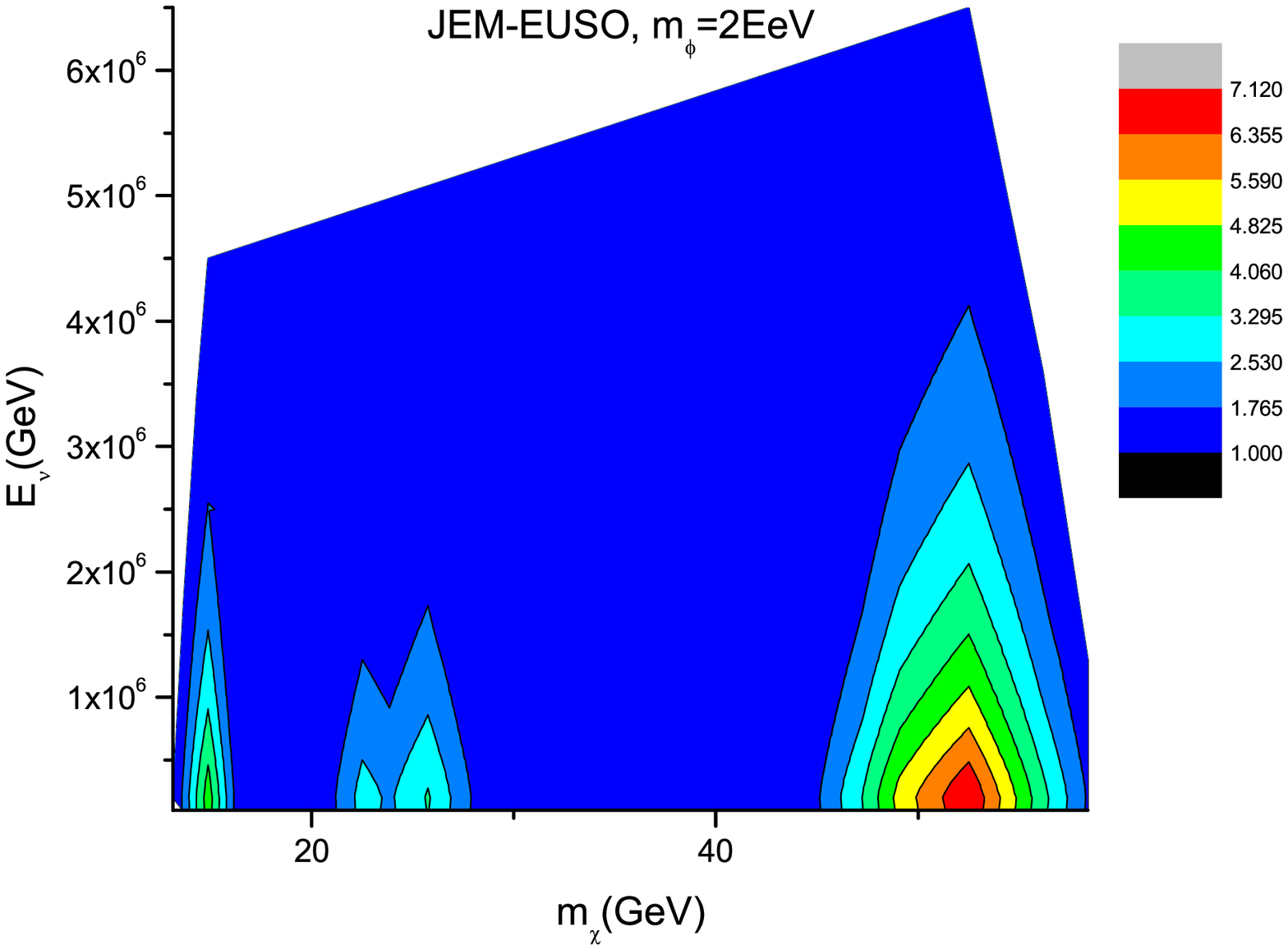}
 \caption{The detected region in the ($m_{\chi}$,$E_{\nu}$) plane is estimated at $m_{phi}=2 EeV$ in JEM-EUSO. $E_{\nu}$ is the energies of neutrinos due to the WIMPs annihilation. The colors represent the number of neutrino signatures from UHE WIMPs in ten years.}
 \label{fig:EUSO_2EeV}
\end{figure}

\begin{figure}
 \centering
 \includegraphics[width=0.9\textwidth]{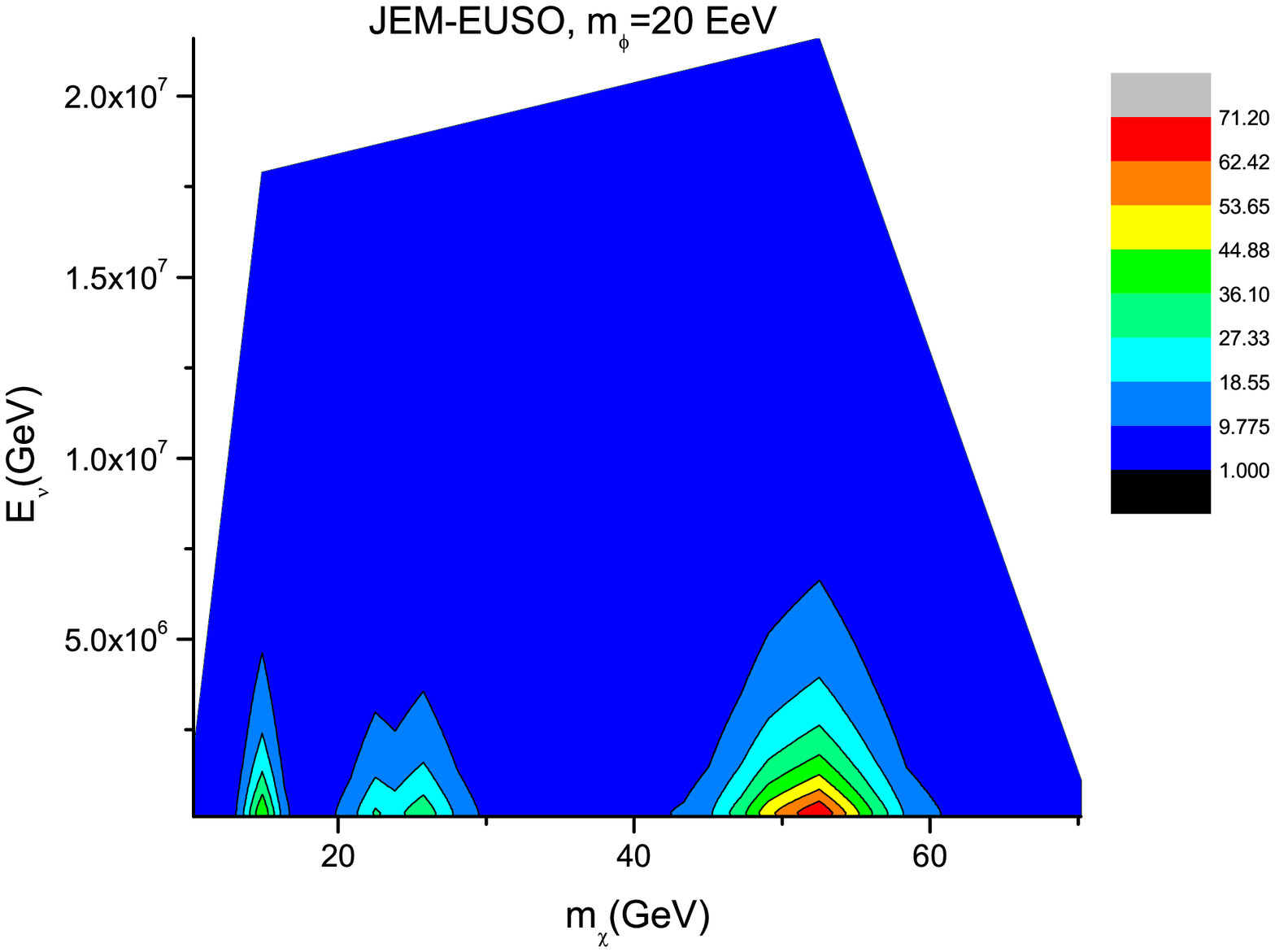}
 \caption{The detected region in the ($m_{\chi}$,$E_{\nu}$) plane is estimated at $m_{phi}=20 EeV$ in JEM-EUSO. $E_{\nu}$ is the energies of neutrinos due to the WIMPs annihilation. The colors represent the number of neutrino signatures from UHE WIMPs in ten years.}
 \label{fig:EUSO_20EeV}
\end{figure}

\begin{figure}
 \centering
 \includegraphics[width=0.9\textwidth]{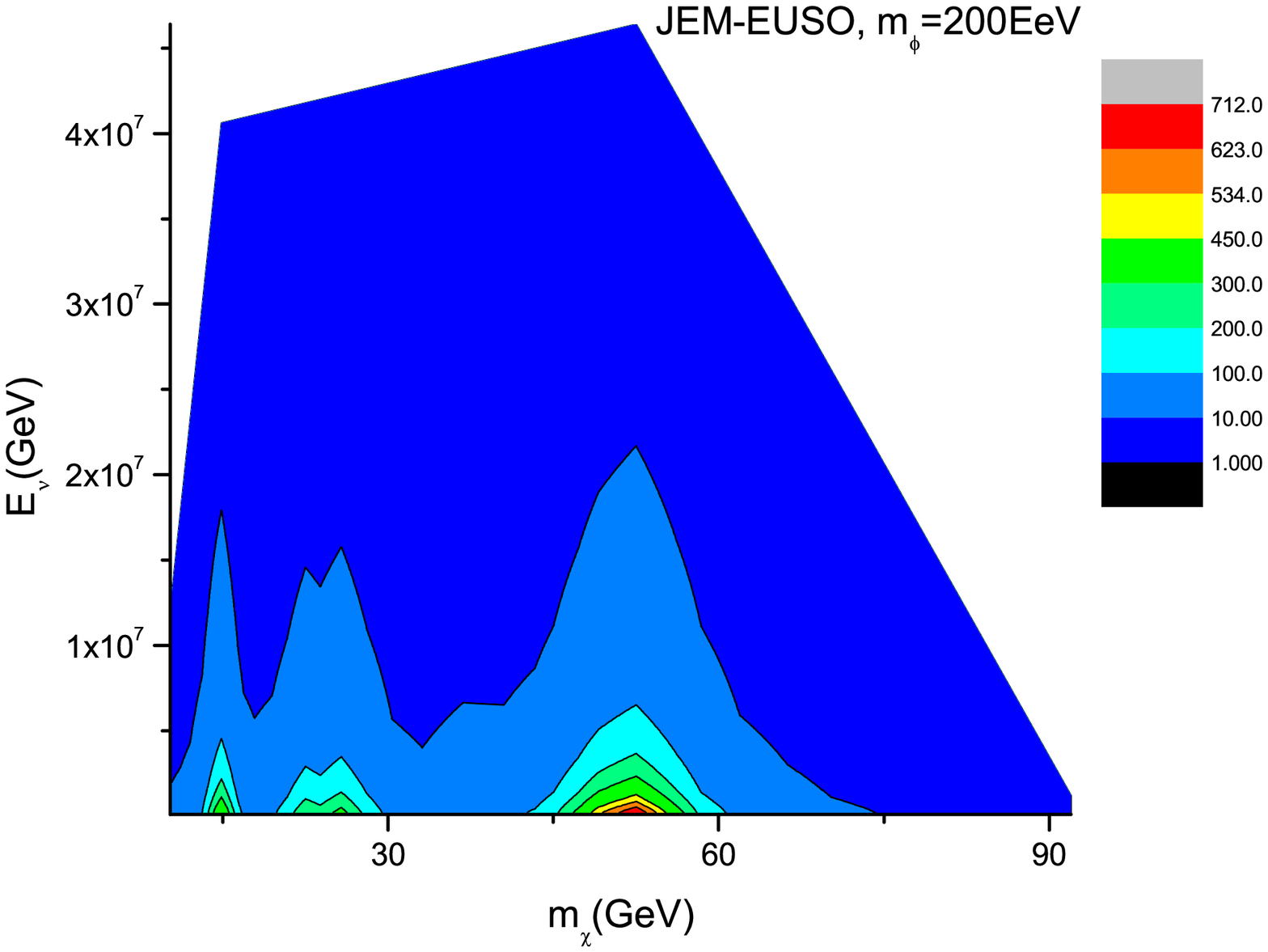}
 \caption{The detected region in the ($m_{\chi}$,$E_{\nu}$) plane is estimated at $m_{phi}=200 EeV$ in JEM-EUSO. $E_{\nu}$ is the energies of neutrinos due to the WIMPs annihilation. The colors represent the number of neutrino signatures from UHE WIMPs in ten years.}
 \label{fig:EUSO_200EeV}
\end{figure}

\begin{figure}
 \centering
 \includegraphics[width=0.9\textwidth]{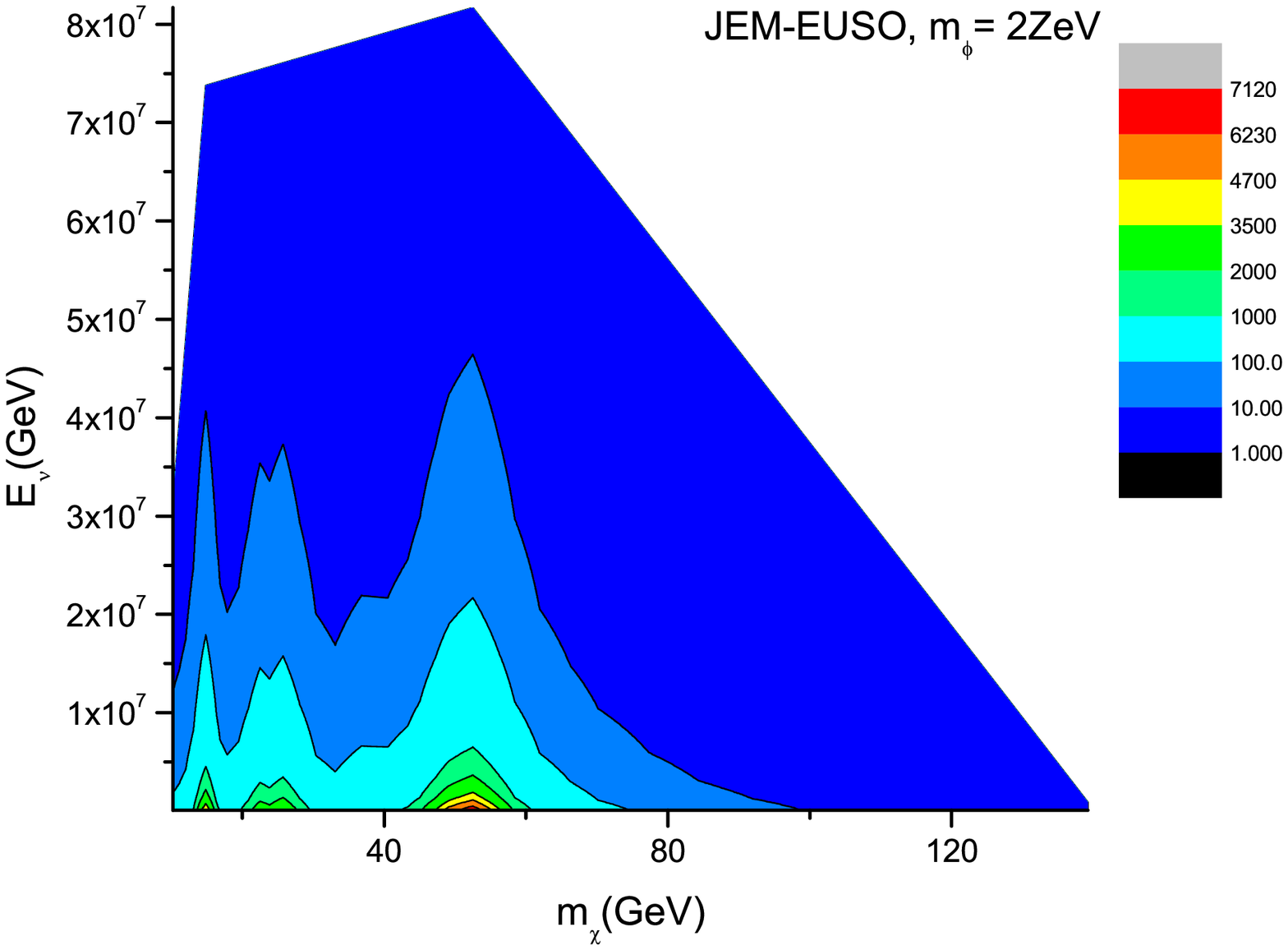}
 \caption{The detected region in the ($m_{\chi}$,$E_{\nu}$) plane is estimated at $m_{phi}=2 ZeV$ in JEM-EUSO. $E_{\nu}$ is the energies of neutrinos due to the WIMPs annihilation. The colors represent the number of neutrino signatures from UHE WIMPs in ten years.}
 \label{fig:EUSO_2ZeV}
\end{figure}

\begin{figure}
 \centering
 \includegraphics[width=0.9\textwidth]{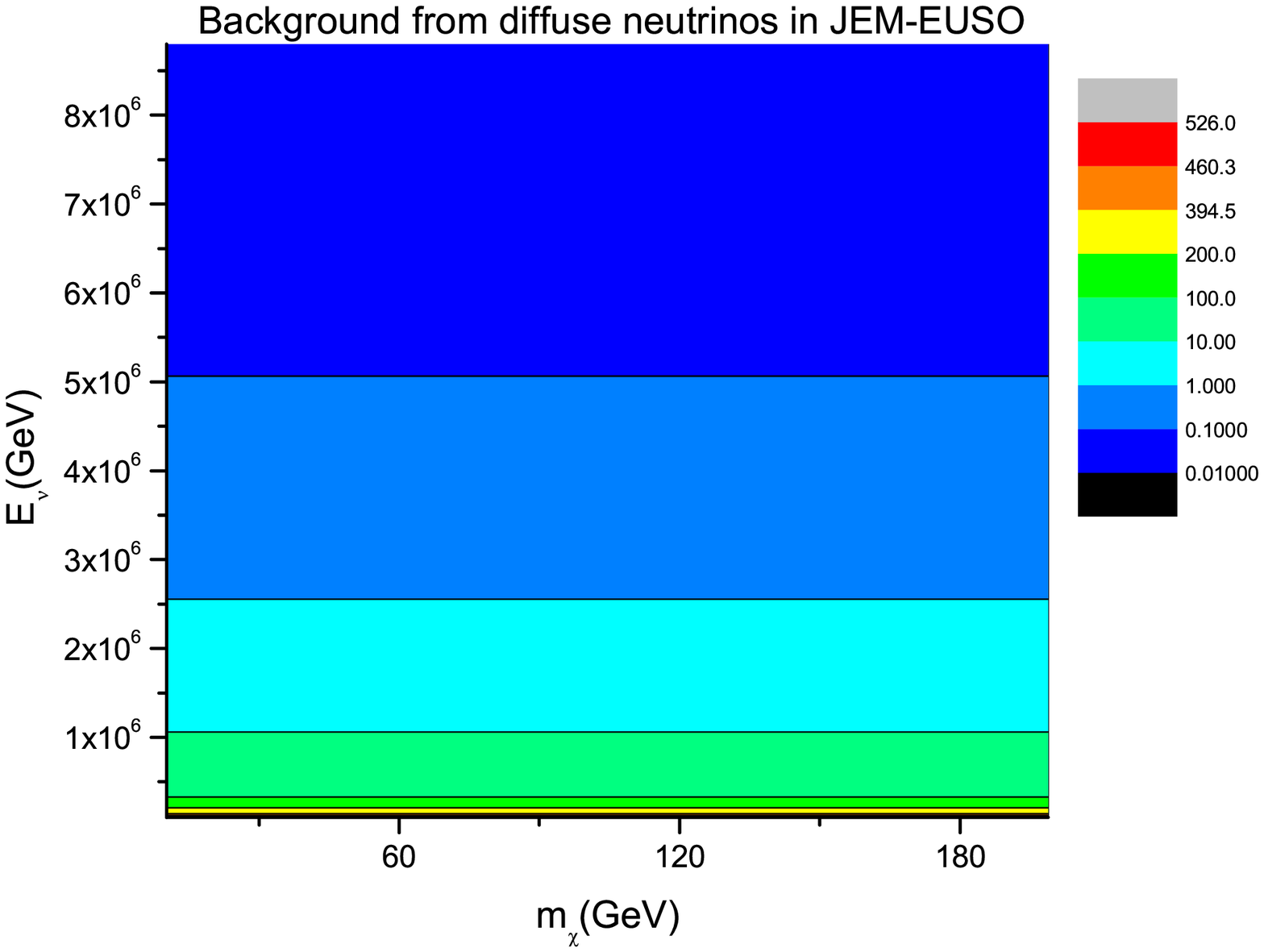}
 \caption{The diffuse neutrino contamination region is roughly estimated in JEM-EUSO. $E_{\nu}$ is the energies of diffuse neutrinos and the colors represent the diffuse neutrino contaminations in ten years.}
 \label{fig:BG_EUSO}
\end{figure}

\begin{figure}
 \centering
 \includegraphics[width=0.9\textwidth]{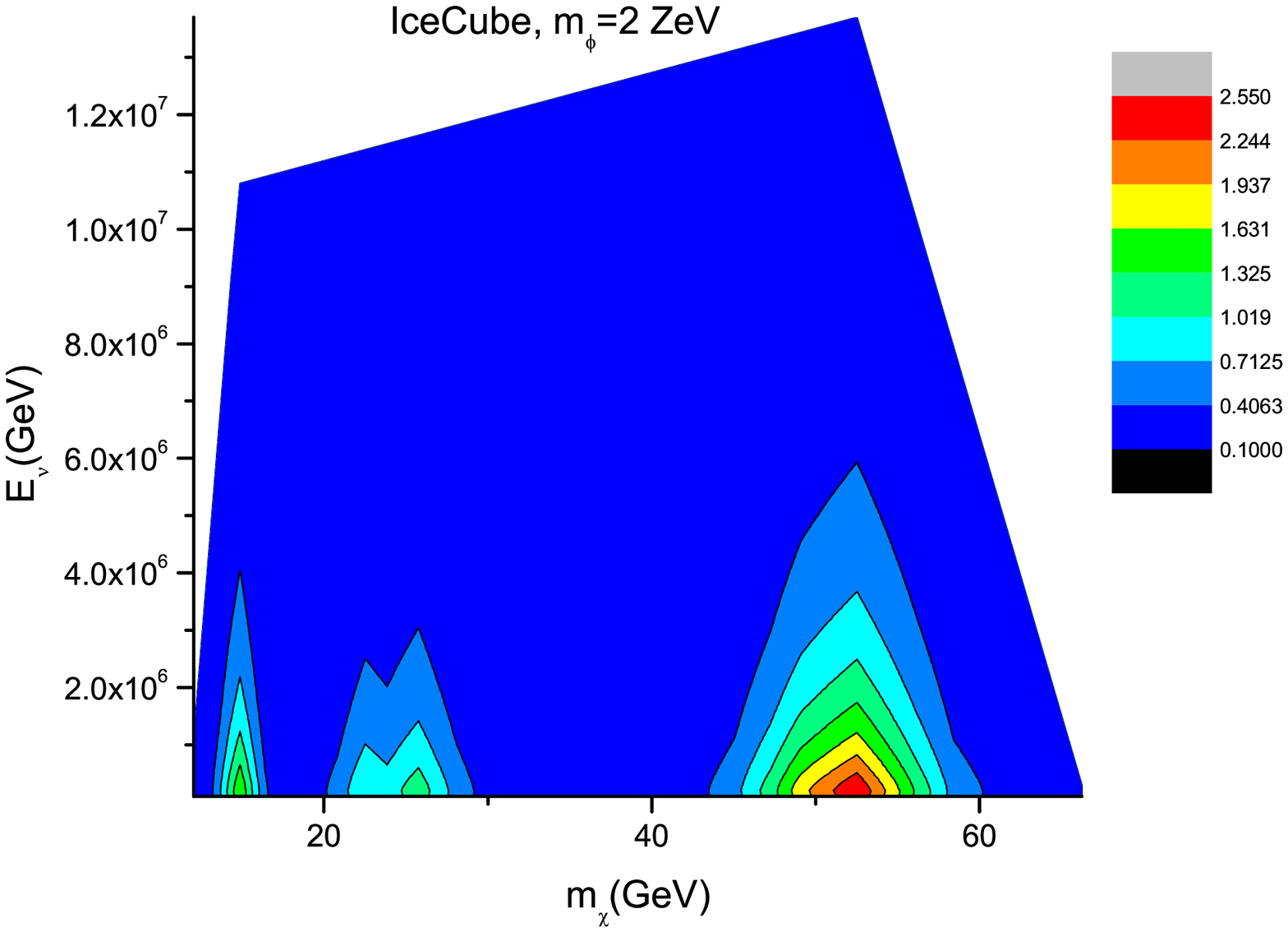}
 \caption{The detected region in the ($m_{\chi}$,$E_{\nu}$) plane is estimated at $m_{phi}=2 ZeV$ in IceCube. $E_{\nu}$ is the energies of neutrinos due to the WIMPs annihilation and the colors represent the number of neutrino signatures from UHE WIMPs in ten years.}
 \label{fig:IceCube_2ZeV}
\end{figure}

\begin{figure}
 \centering
 \includegraphics[width=0.9\textwidth]{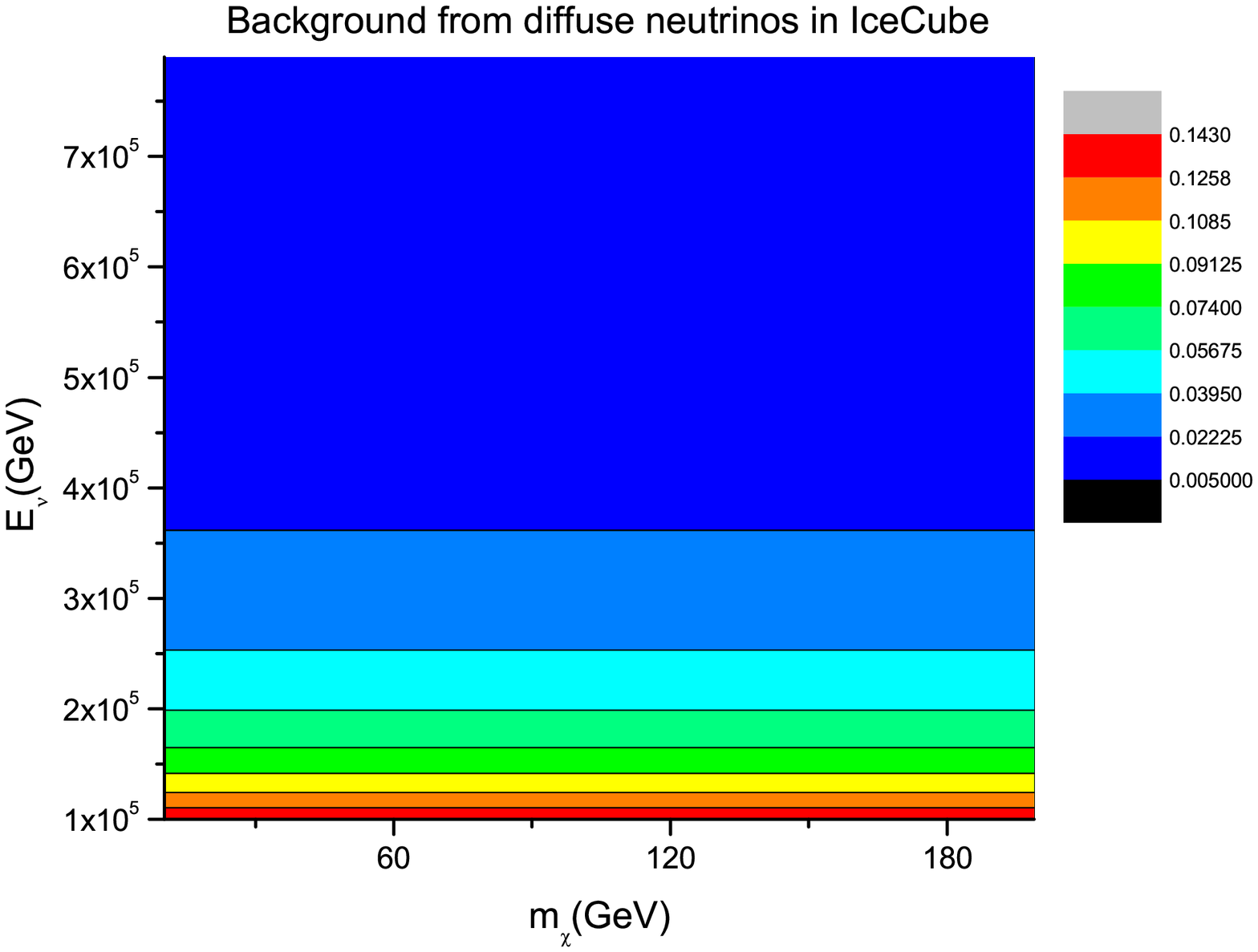}
 \caption{The diffuse neutrino contamination is roughly estimated in IceCube. $E_{\nu}$ is the energies of diffuse neutrinos and the colors represent the diffuse neutrino contaminations in ten years.}
 \label{fig:BG_IceCube}
\end{figure}

\end{document}